\documentclass[twocolumn]{article}
\usepackage[a4paper, total={7in, 8in}]{geometry}

\usepackage{graphicx}      
\usepackage{natbib}        

\usepackage{amsmath,amssymb,amsfonts}
\usepackage{xfrac}
\usepackage{xspace}
\usepackage{bbm}		
\usepackage{array}
\usepackage{algorithm}
\usepackage{algpseudocode}
\usepackage{multirow}
\usepackage{graphicx}
\usepackage{textcomp}
\usepackage{booktabs}
\usepackage{subfig}
\usepackage{footnote}
\usepackage{threeparttable}
\def\BibTeX{{\rm B\kern-.05em{\sc i\kern-.025em b}\kern-.08em
        T\kern-.1667em\lower.7ex\hbox{E}\kern-.125emX}}

\usepackage{balance}
\usepackage{mathtools}
\usepackage{newfloat}
\usepackage{caption}
\usepackage{enumitem}
\usepackage{alphalph}
\usepackage{etoolbox}

\patchcmd{\subequations}{\alph{equation}}{\alphalph{\value{equation}}}{}{}

\usepackage{balance}

\usepackage{tikz}
\usepackage{pgf}
\usepackage{pgfplots}
\pgfplotsset{compat=1.13}
\usepgfplotslibrary{groupplots}

\title{Tracking-in-range Formulations for Numerical Optimal Control\thanks{The work of Alvaro Pons Pelufo, Ying Ju and Giliano-Nicolae Bazili on this topic during their MEng and MSc projects at Imperial College London is gratefully acknowledged.}}

\date{}

\author{Nikilesh Ramesh\thanks{Interdisciplinary Programmes in Engineering, University of Sheffield, S10~2TN Sheffield, UK, email: nramesh2@sheffield.ac.uk}, Eric C.\ Kerrigan\thanks{Department of Electrical \& Electronic Engineering and Department of Aeronautics, Imperial College London, SW7~2AZ London, UK, email: e.kerrigan@imperial.ac.uk}, Yuanbo Nie\thanks{Department of Automatic Control and Systems Engineering, University of Sheffield, S10~2TN Sheffield, UK, email: y.nie@sheffield.ac.uk}}

\begin{document}
\maketitle

\begin{abstract}                
In contrast to set-point tracking which aims to reduce the tracking error between the tracker and the reference, tracking-in-range problems only focus on whether the tracker is within a given range around the reference, making it more suitable for the mission specifications of many practical applications. In this work, we present novel optimal control formulations to solve tracking-in-range problems, for both problems requiring the tracker to be always in range, and problems allowing the tracker to go out of range to yield overall better outcomes. As the problem naturally involves discontinuous functions, we present alternative formulations and regularisation strategies to improve the performance of numerical solvers. The extension to in-range tracking with multiple trackers and in-range tracking in high dimensional space are also discussed and illustrated with numerical examples, demonstrating substantial increases in mission duration in comparison to traditional set-point tracking. 
\end{abstract}


\section{Introduction}
The concept of tracking has been central to control engineering. While the use of set-point tracking (described as error-controlled regulation by~\citet{ashby1956introduction}) has become the de facto standard for such tasks, the notion of `always keeping close to the reference' does not accurately represent the mission specifications of many practical applications. For example, when filming an automotive race event using an unmanned aerial vehicle (UAV) with finite battery resources, the goal may not necessarily be for the target car to be tracked at the centre of the UAV camera until the battery runs out, but instead to keep the car always in the view of the camera (i.e.\ always in range, abbreviated as `a.i.r'), or have the car in the view of the camera for the longest time possible given the duration of the race (i.e. not always in range, abbreviated as `n.a.i.r'). 

The design specifications of such applications can be classified as `in-range tracking' problems. The key characteristic of such problems is that the tracking performance of a dynamic variable is evaluated against a given range $\pm \delta$ around a time-varying reference, with `in-range' being the desired outcome and `out-of-range' being undesirable. 

Importantly,
\begin{itemize}
    \item Inside each sub-category of `in-range' and `out-of-range', there should be no preferences regarding how close the tracking is achieved against the reference point, 
    \item The desire to stay in range should not equate to a requirement for the tracker to always stay in range. In many practical problems, going momentarily out of range may yield overall better outcomes. 
\end{itemize}

\begin{figure}[t!]
 \centering
 \includegraphics[width=0.48\textwidth]{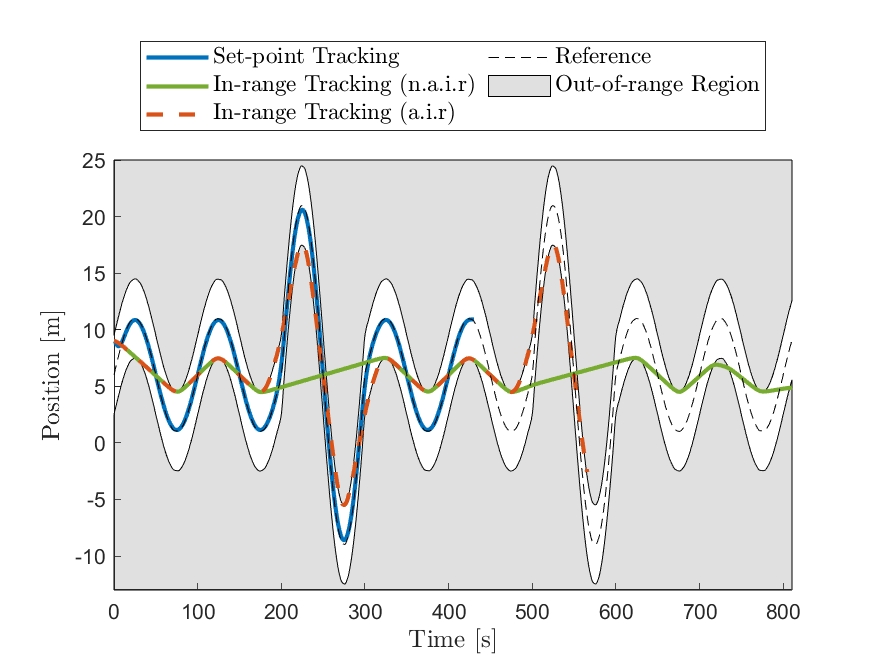}
 \caption{Solutions to a 1-D tracking problem. All solutions have equal energy consumption. `a.i.r' stands for always-in-range and `n.a.i.r' stands for not always-in-range.}
 \label{fig:intro:1d:uav}
\end{figure}

In an optimisation-based control setting, the traditional optimal control problem (OCP) formulations do not represent the nature of `in-range tracking' problems well:
\begin{itemize}
    \item The quadratic regulation cost for set-point tracking, as in~\cite{rawlings2017model}, focuses on the reduction of tracking error magnitude (often subject to a trade-off with control efforts),
    \item For a.i.r problems, formulating an OCP with the range specified as constraints can yield the same result as `in-range tracking' problems. However, for solutions requiring the tracker to go momentarily out of range (i.e.\ n.a.i.r), adapting the constraints as soft constraints ~\citep {kerrigan2000soft} may not be sufficient as such formulation will still discriminate solutions depending on the distance to the range boundaries. 
\end{itemize}
The characteristics mentioned above are visualised in Figure~\ref{fig:intro:1d:uav} for a 1-D tracking problem. For the same energy consumption, in-range tracking with the soft constraint formulation achieves a 31.4\% increase in mission duration in comparison to set-point tracking by keeping the tracker always in range, but not tightly following the target. The in-range tracking formulation that allows the tracker to move out of range can further increase the mission duration by 245s. With an additional in-range time of 54 s, the tracker can now track the target 43.0\% longer than set-point tracking. 

In this paper, we explore suitable OCP formulations that would solve the in-range tracking problems under a trajectory optimisation setting, with the eventual aim of incorporating the design into a nonlinear model predictive control (NMPC) framework. 

Section~\ref{sec:setpoint} and~\ref{sec:in-range} give a formal introduction to tracking problems in optimisation-based control. In Section~\ref{sec:performance}, we propose techniques to improve the numerical performance of the in-range tracking problems and Section~\ref{sec:multiagent} extends the work to multi-agent formulations. Two examples including a multi-tracker and a higher dimensional tracking problems are presented in Section~\ref{sec:numerical}. Finally, Section~\ref{sec:future} highlights the concluding remarks and discussions for future work. 

\section{Set-point Tracking with Optimal Control}
\label{sec:setpoint}

A general OCP can be formulated as such
\begin{subequations}
    \label{eq:to}
    \begin{align}
    \begin{split}
\min_{t_0,t_f,p,x(\cdot),u(\cdot)} \Phi(t_0, t_f, &x(t_0), x(t_f), u(t_0), u(t_f), p) \\ &+ \int_{t_0}^{t_f} L(x(t), u(t),p,t)\: \text{d}t
    \end{split}
    \label{eq:to:cost}
    \end{align}
    \begin{alignat}{2}\text{s.t.\ }
        & f(x(t), \dot{x}(t) , u(t), p, t) = 0, \; \; &&\forall t \in [t_0, t_f] \label{eq:to:dyn} \\
    & g(x(t), \dot{x}(t), u(t), \dot{u}(t),p, t) \leq 0, \; \; &&\forall t \in [t_0, t_f] \label{eq:to:pathconstraints} \\
    & \phi(t_0, t_f, x(t_0), x(t_f), u(t_0), u(t_f), p) &&\leq 0, \label{eq:to:boundaryconds} 
    \end{alignat}
\end{subequations}
with $\Phi$ the Mayer cost, $L$ the Lagrange cost, $f$ the dynamics constraint related to the system differential-algebraic equations, $g$ the inequality path constraints and $\phi$ the boundary constraints. This is an infinite-dimensional optimisation problem with decision variables $x: \mathbb{R} \rightarrow \mathbb{R}^n$ (the states) and $u: \mathbb{R} \rightarrow \mathbb{R}^m$ (the inputs) being functions or trajectories. Static decision variables $p$ can be included as part of the problem formulation as well, and the start time $t_0$ and end time $t_f$ of the problem can also be free. The objective functional~\eqref{eq:to:cost} is often denoted by a single variable $J$ with optimal solution denoted as $J^{\ast}$.

\subsection{Set-point tracking}
In optimisation-based control, set-point tracking is typically achieved through the minimisation of the following quadratic regulation cost
\begin{equation}
\label{eqn: MPCRegCostFunction}
	J = \int^{t_f}_{t_0}e(t)^{\top} Qe(t) + u(t)^{\top} R u(t)\: \text{d}t,
\end{equation} 
with $e(t):=x(t)-x_r(t)$ the instantaneous tracking error between $x$ and reference $x_r$.  $Q \succeq 0 \in \mathbb{R}^{n\times n}$ and $R \succ 0 \in \mathbb{R}^{m\times m}$ are weighting matrices addressing the relative trade-offs between different variables and different terms in the set-point tracking. 

The quadratic regulation cost is typically used under a fixed horizon. As the tracking errors are integrated along the time horizon, having a free $t_f$ would naturally favour solutions with short horizons. To solve problems that require the tracking duration to be maximised, the common approaches are to
\begin{itemize}
    \item iteratively increases $t_f$ that is fixed in the OCP formulation until the OCP becomes infeasible or the tracking performance is no longer acceptable, or
    \item add additional trade-off terms in the objective formulation, e.g.\ $-\omega t_f$ with $\omega$ a trade-off weight.
\end{itemize}

In practice, the idea of having an acceptable range around the reference also often applies to set-point tracking, to evaluate the tracking performance posteriorly or to be implemented as state or output constraints to ensure the tracking performance.  Nevertheless, the preference to reduce tracking error always exists in such formulations.

\section{In-range Tracking} \label{sec:in-range}
In contrast, in-range tracking only focuses on whether the tracker is within a given range $\pm \delta$ around a reference. Within each outcome category of `in-range' and `out-of-range', in-range tracking is invariant to the exact deviation from the reference. Depending on whether `out-of-range' instances are allowed, in-range tracking problems can be tackled differently. 

\subsection{Always-in-range}

Always-in-range (a.i.r) formulation leverages the inequality path constraint~\eqref{eq:to:pathconstraints} to achieve tracking-in-range. This approach strictly confines the trajectory to the desired tracking range $\pm \delta$. We can formulate the constraint as such for a single tracker: 
\begin{equation}
\label{eq:alwaysinrangeconstraint}
     (x(t) - x_r(t))^2 - \delta^2 \le 0, \; \; \forall t \in [t_0, t_f].
\end{equation}

The a.i.r problems are typically formulated with minimum energy or control effort objectives using fixed $t_f$, or with a time duration maximisation objective using free $t_f$. When such path constraints are specified as hard constraints, the state must be initialised within the range and stay in the range. As this may not always be feasible in a real-world engineering situation, practical implementation will typically see~\eqref{eq:alwaysinrangeconstraint} implemented as soft constraints by augmenting the original objective $J$ with additional terms penalising constraint violations, introducing a trade-off between the objective minimisation and constraint satisfaction. 

\subsection{Not always-in-range}

In not always-in-range (n.a.i.r) problems, in-range tracking is achieved via mathematically encoding this tracking behaviour in the objective function along with the original objectives. The stage cost formulation should yield penalties when the tracked variables fall outside the range and yield rewards otherwise, with no preference in the magnitude of deviation from the reference in both cases. 

\begin{figure}[t!]
 \centering
   \includegraphics[width=0.48\textwidth]{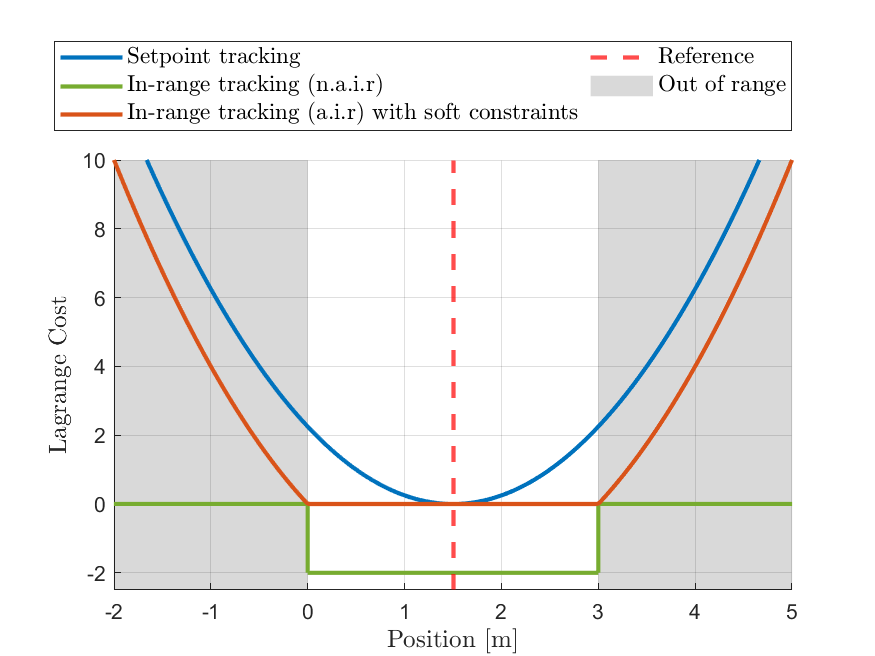}
 \caption{Lagrange cost comparison between different formulations with $x_r = 1.5$ m and $\delta = 1.5$ m}
 \label{fig:tracking:cost}
\end{figure}

A straightforward choice is to use an indicator function of the form for the Lagrange cost
\begin{equation}
\label{eq:OutOfRangeCost}
L_H:=\begin{cases}
            \alpha & \text{for } x_r - \delta \le x \le x_r + \delta \\
            \beta & \text{otherwise}
        \end{cases}
\end{equation}
with $\alpha$ and $\beta$ constants representing respectively the reward for the tracker to be in the range of the target, and the penalty for staying outside the range. A graphical representation of such a function with $\alpha=-2$ and $\beta=0$ is shown in Figure~\ref{fig:tracking:cost}, where a range radius $\delta = 1.5$ m is considered for a target located at a position $x_r = 1.5$ m. The differences between the Lagrange cost for set-point tracking and the soft constraint formulation for a.i.r tracking can also be observed in the figure. 

The desire for the tracker to stay in range can be achieved through the minimisation of the Lagrange cost, i.e.\
\begin{equation}
\label{eq:outofrange:ocp}
    \min_{t_0,t_f,p,x(\cdot),u(\cdot)}  \int_{t_0}^{t_f} L_H(x(t), u(t),p,t)\: \text{d}t
\end{equation}
subject to relevant constraints~\eqref{eq:to:dyn}---\eqref{eq:to:boundaryconds}. If the problem specifications have other original objectives, e.g.\ to minimise energy consumption or control efforts, they need to be augmented to~\eqref{eq:outofrange:ocp} with suitable trade-off weights.

\section{Improving the numerical performance of in-range tracking problems}
\label{sec:performance}

\subsection{Choice of constants in the indicator function}
Due to the integration in~\eqref{eq:to:cost} for the Lagrange cost, the choice of $\alpha$ and $\beta$ will not be shift-invariant, with some design choices better suited for numerical computations than others. Our key observations are
\begin{itemize}
    \item Having $\alpha=0$ and $\beta\ge 0$ should be avoided as different solutions with the tracker staying in the range throughout the time horizons will all have zero costs regardless of the duration of the problem $[t_0, t_f]$, thus worsening the non-uniqueness of the problem solutions.
    \item For problem specifications solely focusing on maximising the in-range time, having $\alpha<0$ and $\beta=0$ is often a good choice as the duration for which the tracker stays outside the range is not directly penalised. This would allow the tracker to comfortably stay outside the range if it is beneficial to increase the overall in-range time. 
\end{itemize}

\subsection{Smooth approximation of the indicator function}
\label{subsec:numerics:smoothapprox}
The design of the Lagrange cost as a discontinuous indicator function will pose many numerical challenges. Therefore it is desirable to use designs of continuous-differentiable expressions to approximate the desired shape of the original function. 

The indicator function~\eqref{eq:OutOfRangeCost} can be viewed as the combination of two Heaviside functions~\citep{abramowitz1968handbook}, for which different smooth approximations exist in the literature, e.g.\ by~\cite{iliev2015approximation}. 

\begin{figure}[t!]
 \centering
 \subfloat[Cost with varying $k_1$ \label{fig:approximation:LQ}]{
   \includegraphics[width=0.48\textwidth]{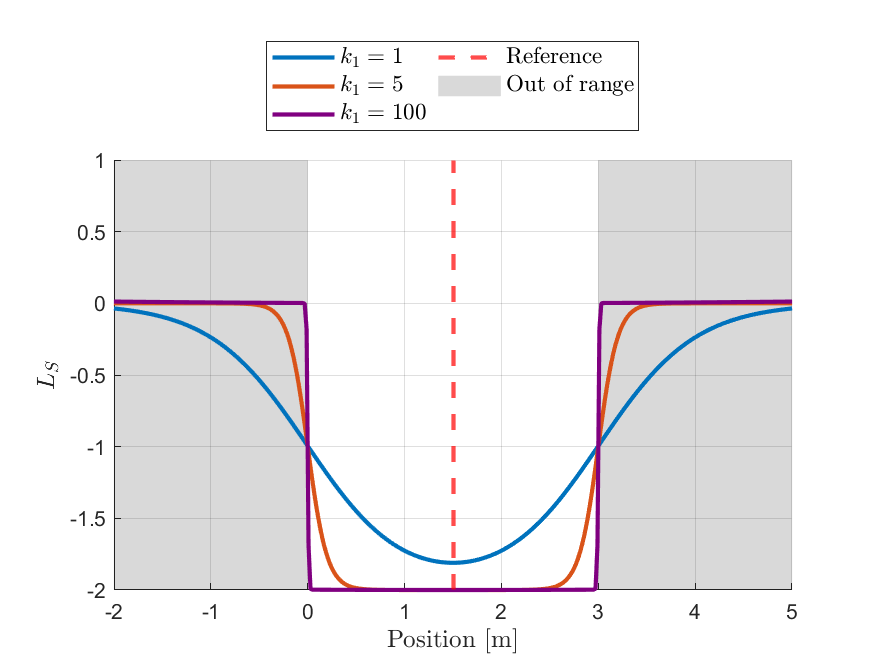}} \\
 \subfloat[Regularised cost with varying $k_2$ and constant $k_1 = 100$\label{fig:approximation:regularised}]{
   \includegraphics[width=0.48\textwidth]{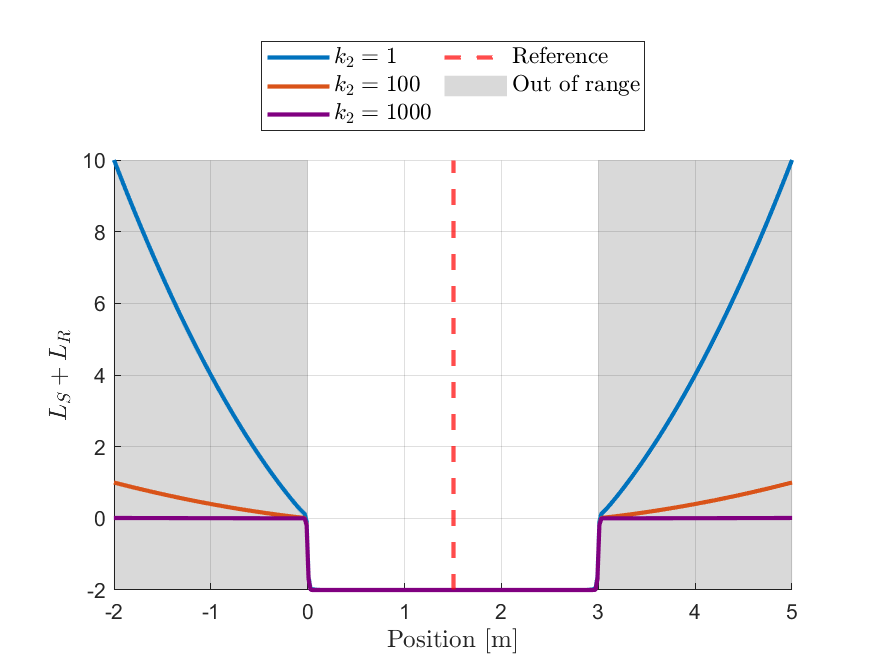}}\\
 \caption{Effect of constants $k_1$ and $k_2$ on the stage cost approximation}
 \label{fig:approximation}
\end{figure}

As an example, with $\alpha=-2$ and $\beta=0$, 
    \begin{align}\label{eq:approximation:LS}
    \begin{split}
    L_H\approx L_S&:=\text{tanh}\left(k_1((x - x_r) - \delta)\right) \\& \quad  + \text{tanh}\left(k_1( -(x - x_r) - \delta)\right) \\
    &=\frac{2}{1+e^{-2k_1((x-x_r)-\delta)}}\\ & \quad +\frac{2}{1+e^{-2k_1(-(x-x_r)-\delta)}}-2.
    \end{split}
\end{align}
With $k_1$ a smooth parameter. Figure~\ref{fig:approximation:LQ} illustrates the effect of choosing $k_1$ on the smooth approximation of $L_H$. 

\subsection{Regularisation}
Despite the smooth approximation, the numerical solution of the in-range tracking OCP could still be challenging due to the lack of gradient information to move out-of-range instances towards in-range. Therefore, we find it practically beneficial to add an additional regularisation term
\begin{equation} \label{eq:notalways:LQ}
    L_R:=\frac{1}{k_2}(x-x_r)^2,
\end{equation}
with regularisation weight $k_2 \to \infty$ for $L_Q \to 0$.

To further reduce the impact of the regularisation term on the trajectory of the tracker when it is in range, we can optionally apply the quadratic regularisation only for the out-of-range regions by having 
\begin{equation} \label{eq:notalways:LR}
    L_R:=\frac{1}{k_2}\max\{ (x-x_r)^2-\delta^2, 0\},
\end{equation}
and employ a smooth approximation of the maximum operator for the computation of the numerical value. For example, \cite{KREISSELMEIER1979113} proposed 
\begin{equation} \label{eq:smoothMax}
\begin{split}
    \max \{s_1,s_2 , ... \; s_n\} & \lessapprox \frac{1}{\rho} \cdot \text{log}(e^{\rho s_1} + e^{\rho s_2} + ... \; e^{\rho s_n}),
\end{split}   
\end{equation}
with the accuracy of the approximation improves as $\rho \to \infty$. Figure~\ref{fig:approximation:regularised} illustrates the effect of varying $k_2$ with $k_1$ and $\rho$ fixed.

Alternatively, we can define
\begin{equation} \label{eq:notalways:LRAlt}
    L_R:=\frac{L_I}{k_2}\left((x-x_r)^2-\delta^2\right),
\end{equation}
with $L_I$ the smooth approximation of an indicator function 
\begin{equation}
\label{eq:IndicatorLI}
L_I \approx \begin{cases}
            0 & \text{for } x_r - \delta \le x \le x_r + \delta \\
            1 & \text{otherwise}
        \end{cases}
\end{equation}
using the same method as in Section~\ref{subsec:numerics:smoothapprox}. 

\subsection{Solution strategy}
With the smooth approximations and the regularisation, the desire for the tracker to stay in-range can be achieved by solving the following OCP
\begin{equation}
\label{eq:outofrange:ocpsmooth}
\min_{t_0,t_f,p,x(\cdot),u(\cdot)}  \int_{t_0}^{t_f} L_S+L_R\: \text{d}t
\end{equation}
subject to relevant constraints~\eqref{eq:to:dyn}---\eqref{eq:to:boundaryconds}.

\begin{figure}[t!]
 \centering
\includegraphics[width=0.48\textwidth]{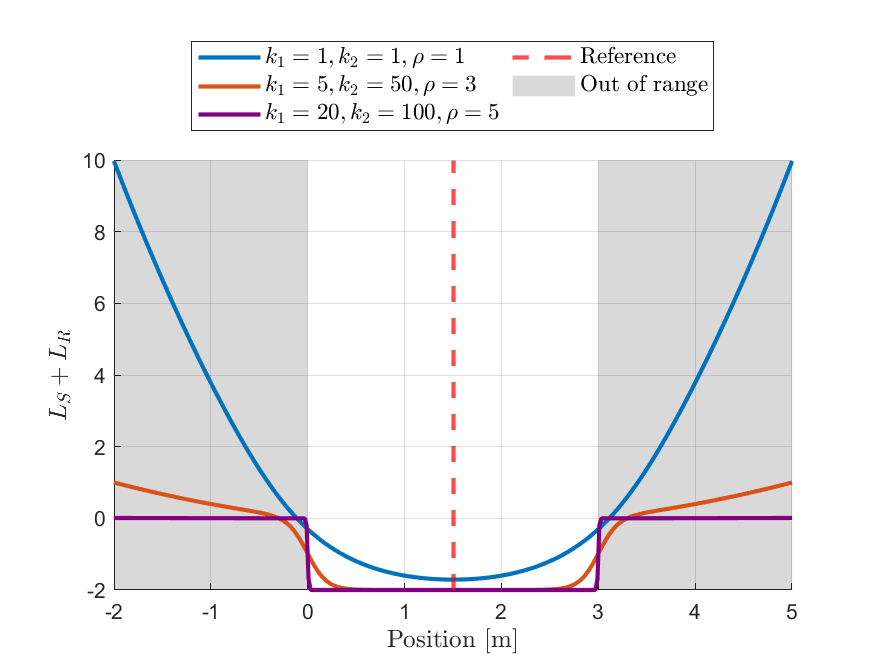}
 \caption{Changes of the Lagrange cost as $k_1$, $k_2$ and $\rho$ increase iteratively}
 \label{fig:tracking:solutionstrategy}
\end{figure}

In practice, it is unlikely that a single choice of $k_1$, $k_2$ and $\rho$ would be sufficient for the solver to efficiently and reliably converge to a solution. Therefore, the OCP may need to be iteratively solved with warm-starting, with $k_1$, $k_2$ and $\rho$ increases after each iteration. Figure~\ref{fig:tracking:solutionstrategy} illustrates the convergence of the Lagrange cost iteratively towards the original cost~\eqref{eq:OutOfRangeCost}, with $\alpha=-2$ and $\beta=0$. 

\section{In-range Tracking with Multiple Trackers} \label{sec:multiagent}

The advantages of in-range tracking are greater in scenarios with multiple trackers, to collaboratively track the reference. The corresponding tracking requirement is typically specified as the desire to have at least one tracker in the range of the target. 

In terms of the OCP formulation, for a.i.r problems, the path constraints~\eqref{eq:alwaysinrangeconstraint} would be reformulated as
\begin{equation}
\label{eq:alwaysinrangeconstraintmulti}
     \min \{(x_i - x_r^2\}_{i=1}^{N} - \delta^2 \le 0.
\end{equation}
for a problem with $N$ agents with $x_i$ the position of the $i$th agent. For n.a.i.r problems, the objective will be based on the minimum value of the Lagrange cost across all $N$ agents, i.e.\
\begin{equation} \label{eq:notalways:L:multi}
    L_S+L_R =  \min \{L_{S_i} + L_{R_i}\}_{i=1}^{N}.
\end{equation}

The minimum operator can be non-smooth and non-differentiable, posing challenges that need to be resolved for practical computation benefits. Here we present two alternatives levering again on the smooth maximum operator or using the complimentary conditions. Here we will only demonstrate the n.a.i.r problems for simplicity, as the treatment for a.i.r problems is analogous. 

\subsection{Using the smooth maximum operator}

As the first step, the minimum operator can be reformulated using the maximum operator as follows
\begin{equation} \label{eq:notalways:L:multimax}
    L_S+L_R =  -\max \{-(L_{S_i} + L_{R_i})\}_{i=1}^{N}.
\end{equation}

Due to particular characteristics of the smooth maximum approximation, direct use of~\eqref{eq:smoothMax} in~\eqref{eq:notalways:L:multimax} could lead to undesirable behaviours. For a given $\rho$, the accuracy of the smooth maximum would decrease when the values of all elements are similar. Added to the fact that the approximation is always larger than the true maximum, the direct use of smooth maximum would yield a lower value for the Lagrange cost when all trackers are in range. Therefore, such implementation would unintentionally introduce biases in the solution, in contradiction to the original specification. 

Practical workarounds are available to mitigate the impact. For example, problems with choices of $\alpha<0$ and $\beta=0$ in~\eqref{eq:OutOfRangeCost} can compute instead
\begin{equation} \label{eq:notalways:L:multimax2}
    L_S+L_R =  \max\{-\gamma \max \{-(L_{S_i} + L_{R_i})\}_{i=1}^{N},\alpha\},
\end{equation}
using the smooth maximum approximation with $\gamma >1$ a sufficiently large positive constant. 

For example, in a two-agent tracking problem with $\alpha=-2$ and $\beta=0$, with at least 1 tracker in the range, \eqref{eq:notalways:L:multimax} should yield a value of $-2$ but the use of smooth maximum approximation with $\rho=1$ would yield $-2.69$ for the instance when both trackers are in range, and yield $-2.13$ when only 1 tracker is in range. When using~\eqref{eq:notalways:L:multimax2} instead with $\gamma=6$, the Lagrange cost with the smooth maximum approximation improves to $-2.00$ in both cases.

\subsection{Using complimentary conditions}
\label{sec:mpcc}

In optimisation literature, the computation of the minimum value operator can be avoided by reformulating the problem into a mathematical program with complementarity conditions (MPCC). \cite{betts} presented the MPCC reformulation for the minimum value operator which can be adapted to the multi-agent in-range problem. 

For a two-agent problem, if we let
\begin{equation}
    \begin{gathered}
        l_1 = (L_{S_1} + L_{R_1}) \\
        l_2 = (L_{S_2} + L_{R_2}) \\
    \end{gathered}
\end{equation}
then 
\begin{equation}
    \min \{L_{S_i} + L_{R_i}\}_{i=1}^{2}=l_1+(l_1-l_2)q,
\end{equation}
if $q$ is chosen to minimise $(l_1-l_2)q$ subject to $-1\le q \le 0$. This problem can be solved directly as a bi-level optimisation problem or reformulated as an MPCC by obtaining the necessary conditions for optimality for the inner optimisation problem and implementing them as constraints in the form of complementarity conditions for the ``outer" optimisation. In this case, we would have
\begin{align*}
    &0 \le \lambda_1 \perp (1+q)\ge 0, \\
    &0 \le \lambda_2 \perp -q\ge 0, 
\end{align*}
with $\perp$ representing the complementarity relationship and $\lambda_1$ the Lagrange multiplier for inequality constraint $-1\le q$ and $\lambda_1$ the Lagrange multiplier for $q \le 0$. 

Although the use of complementary conditions leads to a more accurate representation of the minimum operator, this framework is more difficult to scale with an increasing number of agents, both in terms of the complexity of the problem formulation and the number of decision variables. In addition, solving nonlinear programming problems arising from MPCC can be challenging due to the violation of constraint qualifications~\citep{betts}. As a result, our experience shows that the smooth maximum provides a practically more attractive solution for multiple tracker in-range tracking problems. 

\section{Numerical Examples}
\label{sec:numerical}
To demonstrate the advantages of in-range tracking, two additional example problems are presented to focus on different aspects. All problems are transcribed using the toolbox \texttt{ICLOCS2} \citep{nie2022iclocs2}, and solved with interior point NLP solver \texttt{IPOPT} \citep{wachter2006implementation} to a relative convergence tolerance of $10^{-9}$.

\subsection{Multi-tracker 1D in-range tracking with out-of-range possibilities (n.a.i.r)}

The first example is a 1D in-range tracking problem with two trackers, with the possibility for the tracker to stay out-of-range if desirable. The system dynamics is a double integrator representing simplified UAV dynamics
\begin{equation}
        \Ddot{x}_i = \frac{u_i(t)}{10}, \text{ for } i=1,2 
\end{equation}
together with a simplified energy consumption model with the rate of discharge of the battery state-of-charge ($E_i$ in \%) modelled to have terms dependent on the tracker's velocity $v_i$ and the input $u_i$, and a constant term to account for pseudo-hovering state:
\begin{equation}
        \dot{E}_i = -0.085 - (0.283u_i)^2 - 0.566v_i^2, \text{ for } i=1,2.
\end{equation}
In this example, we will also explore the changes in the solution by using
\begin{align}
    \begin{split}
        \dot{E}_i =& -0.05(0.7+ \text{tanh}(x_i - x_{b}))\\ &- (0.283u_i)^2 - 0.566v_i^2,
    \end{split}  
\end{align}
instead. This dynamics equation includes a slow wireless charging capability when the vehicles are near their base station $x_b$. 

The main objective is to maximise the in-range tracking time for fixed $t_f$ with~\eqref{eq:outofrange:ocpsmooth}, with a secondary objective to minimise the total energy consumption through the maximisation of $E_1(t_f)+E_2(t_f)$. The initial state-of-charge for the batteries, $E_1(t_0)$ and $E_2(t_0)$, are both fixed at 80\%. The trackers are initialised at the base and are required to come back to the base station at $x_{b}=-18$m at the end of the mission with at least 10\% of battery remains. 

\begin{figure}[t!]
 \centering
 \subfloat[In-range tracking (n.a.i.r) without charging\label{fig:1d:doubleuav:Profile1:norest}]{
   \includegraphics[width=0.48\textwidth]{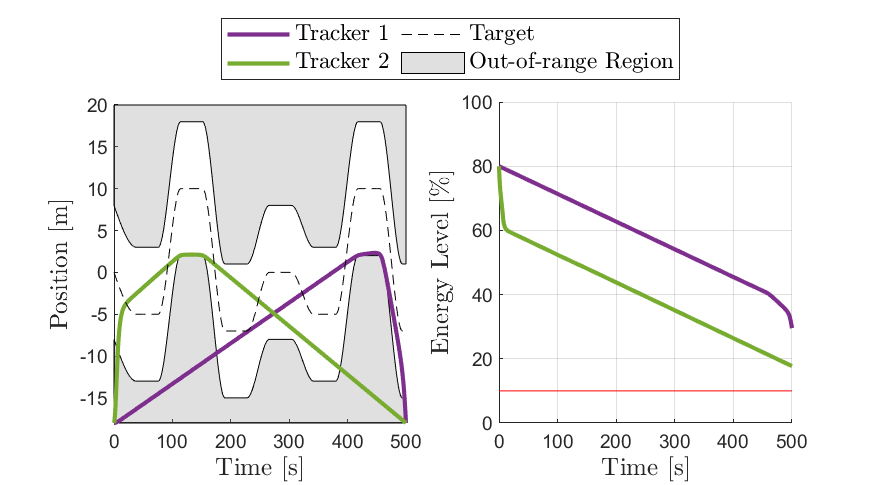}} \\
 \subfloat[In-range tracking (n.a.i.r) with charging\label{fig:1d:doubleuav:Profile1:rest}]{
   \includegraphics[width=0.48\textwidth]{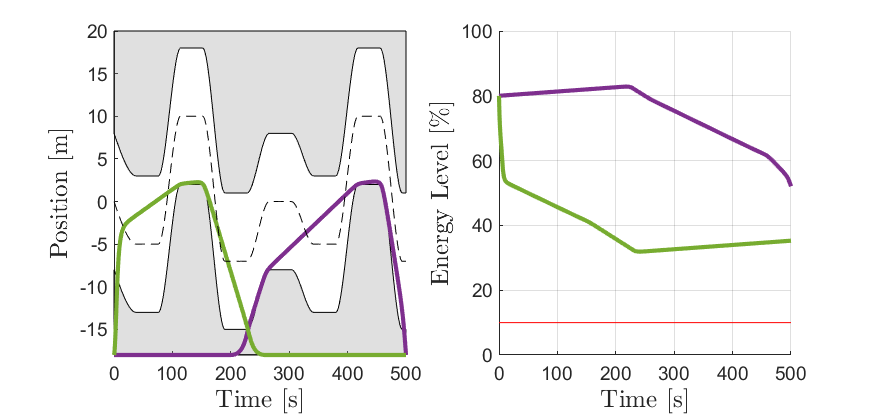}}\\
   \subfloat[Long horizon in-range tracking (n.a.i.r) with charging\label{fig:1d:doubleuav:Profile1:restlong}]{
   \includegraphics[width=0.48\textwidth]{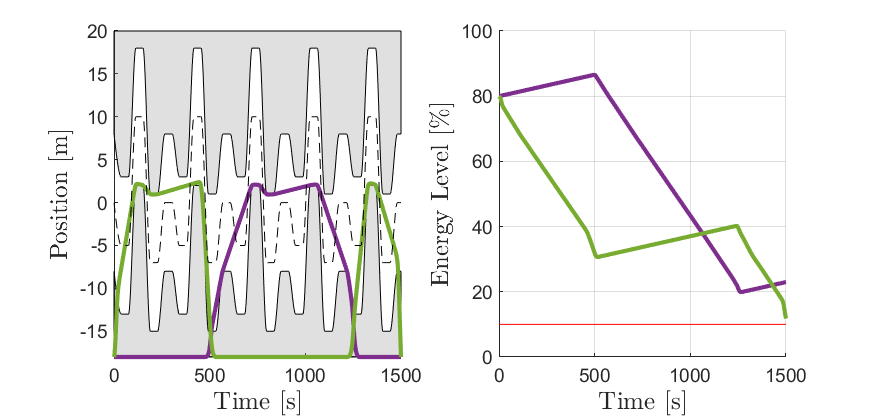}}\\
 \caption{Solutions to multi-tracker 1D in-range tracking problems}
 \label{fig:1d:doubleuav:profile1}
\end{figure}

The problems are solved using the automatic regularisation feature of \texttt{ICLOCS2} with $k_1 = [1, 10, 40]$ and $k_2 = [10^5, 3\times10^5, 10^6]$ respectively for each iteration, and with a choice of $\rho=1$. The solution trajectories are illustrated in Figure~\ref{fig:1d:doubleuav:profile1}.

Comparing Figure~\ref{fig:1d:doubleuav:Profile1:norest} and~\ref{fig:1d:doubleuav:Profile1:rest}, it can be observed that the opportunity to charge at the base has led to substantial changes in the solution: the duration for which both trackers are in-range at the same time has been significantly reduced. With charging and in-range tracking worling together, Figure~\ref{fig:1d:doubleuav:Profile1:restlong} shows that the two trackers can easily keep the target in range for more than 1500 seconds, whereas standard set-point tracking for this problem would typically only allow an endurance of around 350 seconds for each tracker.

\subsection{Single-tracker always-in-range tracking (a.i.r) in 3D}
The simplified model in the previous example has certain characteristics that may give the misconception that in-range tracking always leads to slow moments of the tracker. Therefore we aim to use this second example of 3D in-range tacking with fixed-wing UAV to demonstrate an opposite case. 

\begin{figure}[t!]
 \centering
   \includegraphics[width=0.45\textwidth]{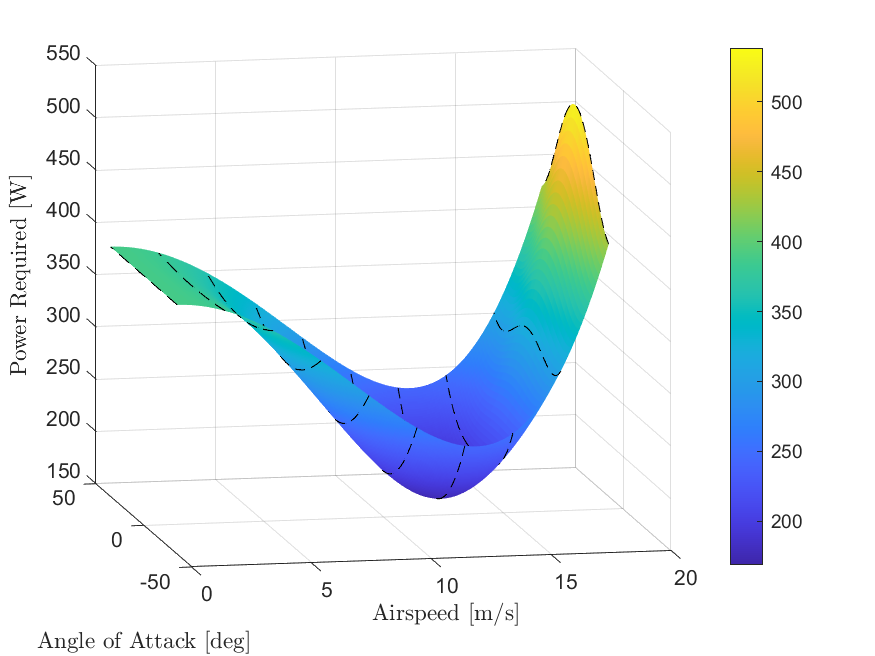}
 \caption{Power required for steady level flight as a function of airspeed and angle of attack}
 \label{fig:3D:powerrequired}
\end{figure}

Based on~\cite{UavPowerConsumption}, the dynamics for the fixed-wing UAV have different characteristics and the power required to maintain steady flight is no longer the lowest at zero airspeed, as shown in Figure~\ref{fig:3D:powerrequired}. In fact, there will be an optimal airspeed to achieve maximum endurance. 

\begin{figure}[t!]
 \centering
   \includegraphics[width=0.48\textwidth]{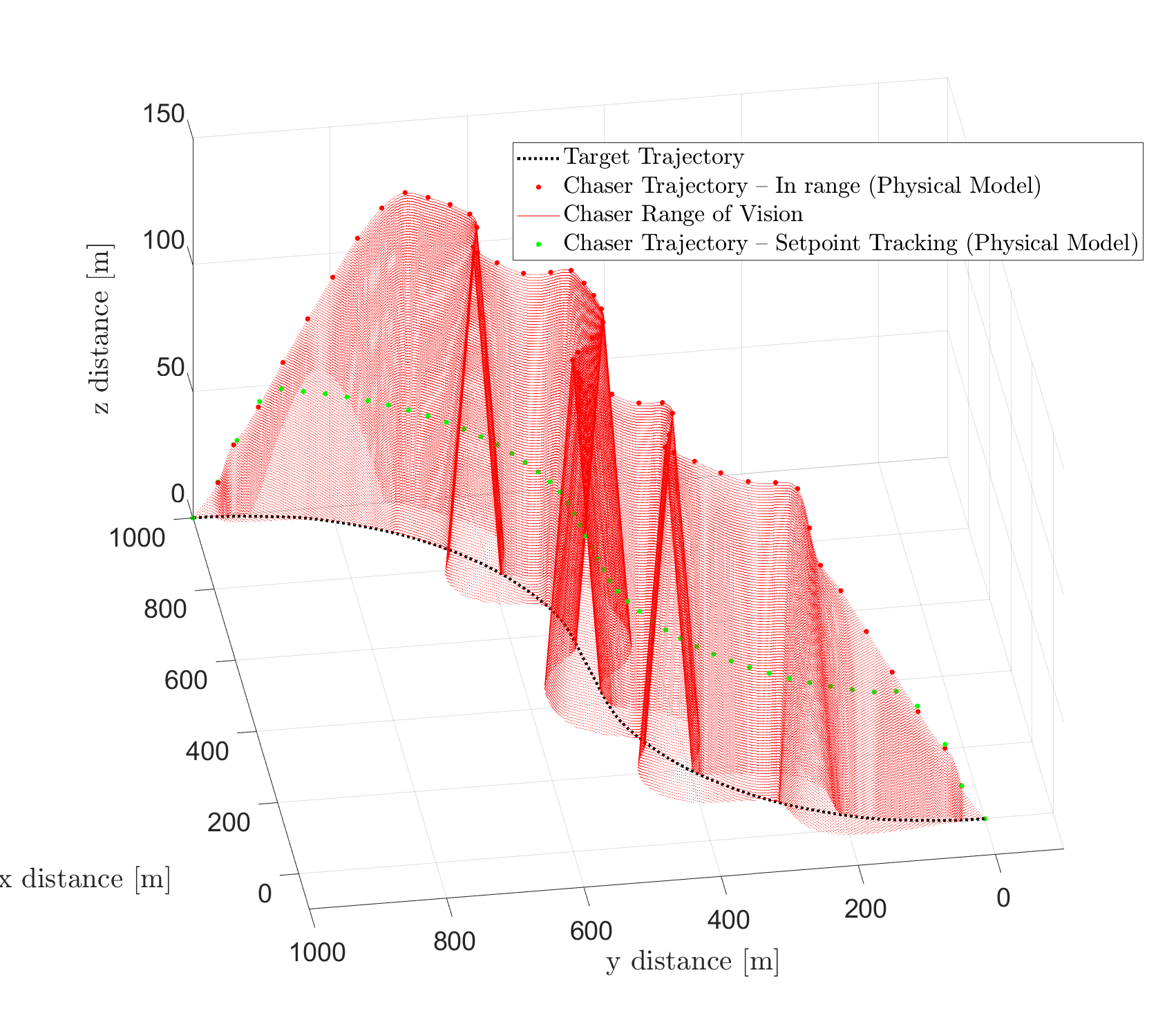}
 \caption{Solution to a 3D always-in-range problem formulation with fixed wing dynamics}
 \label{fig:3D:solution}
\end{figure}

We extended the a.i.r formulation~\eqref{eq:alwaysinrangeconstraint} to a 3D problem, effectively representing an UAV camera having zero-vision at a height of $z = 0$m, and a 40 m radius vision when flying at the ceiling of $z = 130$m. The aim is to keep the target always in the range of the camera's vision. This results in a very interesting solution presented in Figure~\ref{fig:3D:solution}. Distinct from the previous example, fixed-wing aerodynamics give rise to an optimal trajectory where loitering and covering more distance is more efficient. The in-range formulation leverages the optimal cruise conditions to conserve energy, as opposed to set-point tracking which consumes $6.1 \%$ more energy. 

\section{Conclusions and Future Work}
\label{sec:future}

A framework for in-range trajectory optimisation has been presented in this paper. The in-range formulation addresses a number of shortcomings of traditional set-point tracking approaches for some real-world applications. By being invariant to the exact deviation from the reference, the optimal solutions obtained can have improved endurance or reduced energy consumption as shown in the examples. Furthermore, multiple mathematical techniques to improve the performance of these problems such as smooth approximations of indicator and maximum operator functions have been discussed.

Future work would focus on numerically efficient formulations to extend the not-always-in-range problem to higher dimensions, and a detailed analysis of the framework in terms of mathematical properties. Further work will also investigate the solution of in-range tracking problems in closed-loop. 

\bibliography{references}

\end{document}